\documentclass[12pt,preprint]{aastex}
\slugcomment{To appear in The Astrophysical Journal}

\shorttitle{Solar spectrum}
\shortauthors{Livingston et al.}

\begin{document}

\title{Sun-as-a-Star Spectrum Variations 1974-2006}

\author{W. Livingston}
\affil{National Solar Observatory, Tucson, AZ 85726}
\email{wcl@noao.edu}

\author{L. Wallace}
\affil{National Optical Astronomy Observatory, Tucson, AZ 85726}
\email{lwallace@noao.edu}

\author{O.R. White}
\affil{Lazy FW Ranch, Mancos, CO 81328}
\email{orw@hao.ucar.edu}

\and

\author{M.S. Giampapa}
\affil{National Solar Observatory, Tucson, AZ 85726}
\email{giampapa@noao.edu}

\begin{abstract}
We have observed selected Fraunhofer lines, both integrated over the Full Disk and for a small circular region near the center of the solar  disk, on 1215 days for the past 30 years. {\bf Full Disk results:}~{\it Chromosphere:}~ Ca~\footnotesize{II}\normalsize~ K 3933\AA~nicely tracks the 11 year magnetic cycle based on sunspot number with a peak amplitude in central intensity of  $\sim$37\%. The wavelength of the mid-line core absorption feature, called K3, referenced to nearby photospheric Fe, displays an activity cycle variation with an amplitude of  3 m\AA. The separation of the K2 red and blue emission features has increased during the 1976-2006 period of our program. Other chromospheric lines such as He~\footnotesize{I}\normalsize~ 10830\AA, Ca~\footnotesize{II}\normalsize~ 8542\AA, H$\alpha$ , and the CN 3883\AA~bandhead track Ca~\footnotesize{II}\normalsize~ K intensity with lower relative amplitudes. {\it Low photosphere:} Temperature sensitive C\footnotesize{I}\normalsize~ 5380\AA~appears constant in intensity to 0.2\%. {\it High photosphere:} The cores of strong Fe~\footnotesize{I}\normalsize~ lines, Na D1 and D2, and the Mg~\footnotesize{I}\normalsize~ b lines, present a puzzling signal perhaps indicating a role for the 22 y Hale cycle. Solar minimum around 1985 was clearly seen, but the following minimum in 1996 was missing. This anomalous behavior, which is not seen in comparison atmospheric O$_{2}$, requires further observations and theoretical inquiry. 
{\bf Center Disk results:} Both Ca~\footnotesize{II}\normalsize~ K and C~\footnotesize{I}\normalsize~ 5380\AA~intensities are constant, indicating that the basal quiet atmosphere is unaffected by cycle magnetism within our observational error. A lower limit to the Ca~\footnotesize{II}\normalsize~ K central intensity atmosphere is 0.040. This possibly represents conditions as they were during the Maunder Minimum. Converted to the Mt Wilson S-index (H+K index) the Sun Center Disk is at the lower activity limit for solar-type stars. The Wavelength of 
Ca~\footnotesize{II}\normalsize~ K3 varies with the cycle by 6 m\AA, a factor of 2X over the full disk value. This may indicate the predominance of radial motions at Center Disk.  This is not an effect of motions in plages since they are absent at Center Disk. This 11 y variation in the center of chromospheric lines could complicate the radial velocity detection of planets around solar-type stars. An appendix provides instructions for URL access to both the raw and reduced data.

\end{abstract}


\section{Introduction}

The prime instrument of the then new McMath Solar Telescope (first light in 
1962; see \citep{pie64}) was a rapid scan double-pass spectrometer which proved capable of recording the central depths of lines to 0.02\% \citep{bra71}. It occurred to us that, fed by integrated sunlight, the relative strength of a temperature sensitive Fraunhofer line could be used to monitor, from the ground level, photospheric temperature variations and thus the Sun's radiative output. Bob Milkey calculated the contribution function for C~\footnotesize{I}\normalsize~ 5380.323\AA~ and showed its origin to be within 25 km, or practically coincident, with the $\tau$ = 1.0 continuum. Temperature sensitivity was estimated at $\Delta r/r_{0} = .002~ deg^{-1}$~ \citep{liv77}. Unlike typical Fe lines, which strengthen toward the limb, C~\footnotesize{I}\normalsize~ $\lambda$5380 weakens and finally disappears there, in support of Milkey's modeling. With a suitable optical train of flat mirrors in place for a feed, a 3 month run on the carbon line yielded a variance of 0.85 K rms (or 0.06\% in the solar constant).

One of us (ORW) pointed out that the UV component of solar irradiance is also important to the Earth and that chromospheric Ca~\footnotesize{II}\normalsize~ K intensity at 3932\AA~ could serve as a UV surrogate of the magnetic Sun. An observational program involving Ca~\footnotesize{II}\normalsize~ H and K, together with nearby continuum windows, was devised. Regular C~\footnotesize{I}\normalsize~ 5380 and Ca~\footnotesize{II}\normalsize~ H and K observations began. Since 1974 we have averaged 3 observations per month and these continue up to the present.
 
The Ca~\footnotesize{II}\normalsize~ H and K observations were successful from the beginning. Similar observations also were begun at Sacramento Peak Observatory in 1976, and the two data sets may in principle be merged. We attribute the success
of these programs to the large cycle modulation in the H and K lines. A number 
of papers have been published over the years ( see \citet{whi98} and references therein), and the Ca~\footnotesize{II}\normalsize~ H and K index (a 1\AA~integral of the central intensity over the K232 features at line core; a comparable parameter to that used by stellar observers \citep{hal95}) is established as a reliable index of chromospheric activity. We also compute a 0.5\AA~ Ca~\footnotesize{II}\normalsize~ K index by integrating the line profile between the K3V and K3R minima.  This index isolates the chromospheric emission and shows a larger amplitude variability than the 1\AA~index.  Two other Ca~\footnotesize{II}\normalsize~ K features, the wavelength of K3, and the spacing of the K2V and K2R components, are found to display interesting variations over the 11 yr cycle. Ca~\footnotesize{II}\normalsize~ K3 position varies in phase with the activity cycle. Also, over our observing time, the separation of the K2 emission peaks has steadily widened. Other chromospheric lines such as He~\footnotesize{I}\normalsize~ 10830\AA, H$\alpha$, Ca~\footnotesize{II}\normalsize~ 8542\AA, and the CN bandhead at 3883\AA~were added to the program and found to vary in strength like Ca~\footnotesize{II}\normalsize~ K.

The goal of inferring temperature change in the photosphere proved to be more challenging. In part this is because the expected line depth variations are on the order of 1\% or less. One of us (LW) found that many lines were blended 
by weak water vapor lines and an empirical correction was devised for such blends based on near simultaneous measures of Earth water lines near He~\footnotesize{I}\normalsize~ 10830 (see Section 6). For early data before helium was observed we learned to use radiosonde observations obtained from the local weather station. In general, our observational methods proved not to be as completely free of instrumental effects as we initially thought.  In the case of a  photospheric line such as C~\footnotesize{I}\normalsize~ 5380\AA,  the variance in central depth can be as small as 0.05\% on a given day.  But day to day differences are much larger, and this is unlikely due to the Sun. One variable is spectrograph alignment. In a number of early papers we report on the Sun's temperature variability \citep{liv77}\citep{liv82}\citep{liv88} and the reader should realize that these results are no longer valid.  We discuss these problems in the next section. Within our capability to measure it using the C~\footnotesize{I}\normalsize~ 5380\AA~line the global (Full Disk) and basal (Center Disk) photospheric temperature is constant over the activity cycles 21, 22, and 23.

Some of our results, particularly with Ca~\footnotesize{II}\normalsize~ H and 
K, will be of interest to stellar astronomers.  One of us (MSG) therefore 
highlights these facets in the stellar context.

We recognize that our participation in this research is ending. In this paper 
we aim to summarize what has been learned as objectively and succinctly as 
possible. This unique data-set can form the basis of many other interesting
studies by members of the scientific community.  In view of this, we provide
an appendix of information on how the reader can access both the raw and 
reduced archives. This synoptic program of Ca II observations, as well as other
key spectral diagnostics in the solar spectrum, is expected to be continued with the 
Integrated Solar Spectrometer on the SOLIS Instrument at Kitt 
Peak \citep{kel03}.

\section{Observations and their deficiencies}

We discuss first the spectrograph and then the Fourier Transform Spectrometer (FTS).

Fig. 1 shows representative spectra of the bands we have studied in detail for photospheric and chromospheric variability.  These examples illustrate the quality of the measured spectra.  The intensity measures given in this paper are relative to a continuum intensity when a flat continuum was observed.  Since the 
7.5 \AA~scans in the Ca~\footnotesize{II}\normalsize~ K line do not include 
a continuum, the intensity reference was taken at 3935.6 \AA~ in the red wing of the K line.  This reference was taken to be 0.1762 of the nearby continuum 
value at 3953 \AA~ between the K and H lines \citep{whi81}.

Note that in this paper we use the term `central depth' to be the relative line core intensity difference below the local continuum in weak to moderate strength lines. The term `central intensity' in well resolved strong lines like Ca~\footnotesize{II}\normalsize~ K is the minimum intensity in the line core above the zero level. Obviously, central intensity = 1~$-$~central depth.  We present equivalent width variations in He~\footnotesize{I}\normalsize~ 10830\AA~and C~\footnotesize{I}\normalsize~ 5380\AA~as a measure of the total line strength. We also arrange all plots of line strength so that maximum activity signals are upwards (positive) for ease of interpretation (even if this means inverting the sign of the ordinate).
 
Full Disk, or integrated light, is achieved by replacing the telescope concave image forming mirror with a flat and allowing the spectrograph entrance slit to form a `pinhole' image of the Sun on the grating. While this arrangement is imperfect it would be time invariant providing all other optical elements are fixed. This last condition was violated by a grating size change in 1992 and because of frequent collimation adjustments thereafter.

The photometry system has been described by \citet{bra71}. This double-pass system involves rapid scanning of the grating with signal detection behind the exit slit using a photomultiplier or silicon diode. Double-pass means that spectrograph scattered light and electronic off-sets are fully compensated. Variance in the spectra can arise from several factors: photon noise, scintillation, spectrograph seeing, sky transparency (clouds), and grating size (or collimation) effects. {\it Photon noise:} is rendered negligible (0.02\%) by averaging 30 scans and making 6 sets of runs for each `observation'. {\it Scintillation noise:} is a factor because the telescope effective aperture is only a few mm -the entrance slit size. Scintillation is also suppressed by the 30 scan, 6 sets of runs, routine. {\it Spectrograph seeing:} can be a problem especially in the winter when the top of the vertical spectrograph tank becomes colder than its bottom and internal convection sets in. We noticed that broad lines like H$\alpha$ show much less noise under these conditions than sharp photospheric lines. Again this is a random noise which should average out.

 {\it Grating size and collimation error} are more serious problems. The original grating (25 $\times$ 15 cm) was replaced by a larger one (42 $\times$ 32 cm) in 1992. The 15 cm width of the original meant that part of the solar limb fell outside the grating; whereas, the full image was sampled by the new grating. Because most Fraunhofer lines strengthen near the limb we were now observing a different star with the new grating, and a step change will take place in the temporal record of line intensities. One would expect the variance to be reduced with the new grating because the entire disk lies within the larger ruled area. In fact it went up. This could be due to the presence now of two gratings for different experiments, i.e. an IR and a Visible one. The need to recollimate with its unavoidable errors occurred because of almost daily grating changes. We emphasize that Ca~\footnotesize{II}\normalsize~ K displays only a small center-to-limb variation and so was less affected by the change to the larger grating in 1992. A general discussion of observational error is given in \citet{liv82}.

Operations at the FTS present different problems. Collimation is accurate enough and reproducible: a laser beam is sent through the FTS optics and out the entrance hole to pass through the entire telescope in autocollimation mode and then falls back onto this hole. The beam splitter, however, was rather non-uniform in reflectivity so that the sharply focused solar pinhole image, which rotates at a diurnal rate on the collimator due to the heliostat feed, thereby continually weights slightly different parts of the Sun.

Another noise source is scintillation. Telescope aperture is again small, only 5 mm, this being the size of the entrance hole to the FTS. The interferometric central fringe, which sets the DC level of the spectrum, is sampled for only a fraction of a second for each scan. We suspect this is the reason that different FTS records show significant differences in line depth on the same day. These differences can be much greater than would be expected from the apparent high S/N in the FTS spectra.

The FTS does have certain important advantages over the spectrograph. Lines can be selected for study, ex post facto, anywhere within the observed bandwidth of 5000-6300\AA. The wavelength scale of FTS spectra is very accurate.  And the FTS optical system has not changed appreciably over the observational time span 1980-2006. Unfortunately the aging FTS has not been available as often as we would like during the past few years.

\section{Results for Chromospheric Lines}
Fig. 2 compares sunspot number (World Data Center, Belgium) with total unsigned magnetic flux and with the spectral strength of He~\footnotesize{I}\normalsize~ 10830\AA, both of the latter observed almost daily until recently with the Kitt Peak Vacuum Telescope. These KPVT observations are area scans which can be summed to yield the average magnetic field flux and He~\footnotesize{I}\normalsize~equivalent width over the entire disk. Sunspot numbers are the daily values and, of course, this time series is uninterrupted. The total field and He~\footnotesize{I}\normalsize~ equivalent width records are smoothed with a 4 point running mean and contain unmarked time gaps because of clouds and equipment down time. 

Fig. 3 compares the Vacuum Telescope record with our far less frequent McMath-Pierce spectrograph data for He~\footnotesize{I}\normalsize~ equivalent width and the Ca~\footnotesize{II}\normalsize~ K index. The latter two are shown without any running means and include all observations. Some detailed differences in the time series can be seen, but we have not attempted to study these. The main point here is that chromospheric lines respond to the solar activity cycle and its magnetic signature on the solar disk.

The Ca~\footnotesize{II}\normalsize~ H and K reduction program measures and archives a number of spectral features described by White and Livingston (1981). The wavelength of K3 is computed using a parabolic fit to Fe 3932.637\AA~as a reference. Fig. 4 shows a temporal record of the K3 wavelength, both Full Disk and Center Disk, after the application of a 40-point running mean. (\citet{don88} proposed the use of such a filter to reduce the effect of rotational modulation). The sense is a 3 m\AA~blue shift at solar maximum full disk. One might suspect the effect is caused by the cycle plage modulation of the K2V feature. Plage modulation is ruled out by two further pieces of evidence. First, the amplitude of the wavelength shift increases to 6 m\AA~ at Center Disk where little or no plage occurs. This also suggests the motion is mainly radial at Center Disk. Second, in Fig. 5 we show single day observations of the K232 features at minimum and maximum epochs full disk. If K2V were the cause of the resulting wavelength shift, at maximum the shift would be to the red. Instead the shift is to the blue. This figure also suggests that it is the K3 component that is moving, and not K2V-K2R; the upper chromosphere as indicated by K3 is shifting and not the reference Fe line. Admittedly, the K2R feature seems to shift slightly at minimum in this data set; whereas, K2V does not. But these are single day, single record observations containing some noise. 

Another Ca~\footnotesize{II}\normalsize~ anomaly is the wavelength separation of the K2V and K2R components, Fig. 6. Particularly at solar minimum, there 
seems to be a secular increase in this value. This is the presumed basis for 
the change of yet another parameter: the Wilson-Bappu 
effect \citep{whi81}, also shown in Fig. 6. The W-B effect is defined as 
the log of the separation of the half intensity points between the K2 peak and 
the K1 minimum on both sides of the line. This measurement is intended 
to mimic the classical visual micrometer estimates from stellar spectra.

In Fig. 7 we see that the cycle minimum value for the Full Disk Ca~\footnotesize{II}\normalsize~ K index is 0.086, for Center Disk Ca~\footnotesize{II}\normalsize~ K index the mean is 0.073, and for Ca~\footnotesize{II}\normalsize~ K3 it is 0.055. See Fig. 1(b) for the definition of these parameters. These values at Center Disk are with the 2 arc-min spatial averaging lens in place. We can enquire what the value would be for a quiet region at higher resolution.

In Fig. 8 we show such an observation following a suggestion of N.R. Sheeley Jr. We scanned across a quiet region near solar minimum at a resolution of about 2 arc-sec. The minimum value for the Ca~\footnotesize{II}\normalsize~ K3 intensity is about 0.040. This intensity represents the lower limit of the quiet atmosphere at this epoch in the activity cycle. A similar result was reported by \citet{sku84} except they used the Ca~\footnotesize{II}\normalsize~ K index. On the same day and conditions the above was observed we also made a 2 arc-sec scan of C~\footnotesize{I}\normalsize~ 5380. We might expect granular modulation of this line but the spatial resolution was insufficient.

The behavior at Center Disk for Ca~\footnotesize{II}\normalsize~ K3 intensity is also given in Fig. 8. The Ca~\footnotesize{II}\normalsize~ K profile Center Disk is essentially the same as Full Disk at minimum phase (Fig.1). Initially, in 1975-1982, the observations were always taken at the geometric center of the disk. This does not mean that we are necessarily on the Sun's equator since this position varies $\pm$7 degrees due to the inclination of the solar rotation axis to the plane of the ecliptic. Starting in 1983, and continuing up to the present, if any K2 emission indicating magnetic activity (plage or network) is present near disk center, the image was displaced to a nearby quiet area showing the weakest K2 emission. Once this procedure was adopted, the Ca~\footnotesize{II}\normalsize~ K Center Disk temporal record no longer showed any evidence of the 11 y cycle. Note also that the grating change in 1992 was imperceptible in Center Disk Ca~\footnotesize{II}\normalsize~ K. This is because the grating was always fully filled (unlike for the Full Disk data). The remaining short term variations in the Center Disk 
measurements may arise from the transient presence of ephemeral 
regions \citep{har93}.  A comparison of the Center Disk 
Ca~\footnotesize{II}\normalsize~ K index with that for the Full Disk is 
encapsulated in Fig. 9 where the histograms are the distributions of the index
for the Center Disk and Full Disk, respectively.  The two 
distributions are clearly segregated with negligible overlap at the 
minimum values.

Finally, we computed the rms deviation in the annual means for the Full Disk Ca~\footnotesize{II}\normalsize~ 
K index. These results are seen in Fig. 10 where the solar cycle variation is 
readily apparent. We also note, however, that there appears to be a secular
decline in the rms during the last three cycles. We speculate that this 
could indicate a long-term increase in the relative homogeneity of regions 
that are the principal contributors to the Full Disk K index. 

We may comment that no cycle variation is seen in the strength of the very weak rare earth emission lines of Ce~\footnotesize{II}\normalsize~3967.043 and Nd~\footnotesize{II}\normalsize~3934.798 found in the wings of H and K, see Fig. 1 \citep{eng73}. Time series analysis shows the relative intensity of Ce~\footnotesize{II}\normalsize~ is (0.005$\pm$0.0006)+0.00016y, and Nd~\footnotesize{II}\normalsize~ is (0.1$\pm$0.003)$-$(6.8e-06)y, where y=year$-$1900. These lines are only visible Full Disk and not at Center Disk.

Following the grating change in 1992 we ceased to measure the continuum near the H and K lines. Using the \citet{hou70} high points by means of special scans at 3910, 3953, and 4020 \AA, the purpose was to define the amount the Ca~\footnotesize{II}\normalsize~ K wings vary over the cycle. This quantity was reported in \citet{whi81}. Because the color distribution from the new grating was quite different after the grating change, we chose to simplify the observing program and discontinue these continuum reference measurements.
 
In summary the peak-to-peak Full Disk cycle variations found in the K line are: 1 \AA~index~=~25\% (Fig. 7); 0.5 \AA~index~=~35\%; K3 intensity~=~37\% (Fig.8).
In Figs. 11 and 12 we add the data for chromospheric He~\footnotesize{I}\normalsize~ 10830\AA, Ca~\footnotesize{II}\normalsize~ 8542\AA, H$\alpha$, and the CN bandhead at 3883\AA. The CN bandhead represents the low chromosphere where magnetic features are more compact. Title's (1966) collected spectroheliograms allows us to judge that Ca~\footnotesize{II}\normalsize~ K, Ca~\footnotesize{II}\normalsize~ 8542\AA, and H$\alpha$ display plage and faculae with comparable but decreasing contrast in that order. CN 3883, when the seeing is good, maps faculae particularly well (\citet{bai69}, \citet{she69}). He~\footnotesize{I}\normalsize~ 10830\AA~spectroheliograms map the highest levels of the chromosphere \citep{gio77}. Ca~\footnotesize{II}\normalsize~ 8542\AA~spectroheliograms map the lower levels of the chromosphere  nearer the limb \citep{jon83}.

\subsection{Stellar Perspectives}

In this section we discuss some aspects of our Sun-as-a-star 
results in the context of stellar observations.  In particular, we 
examine the behavior of the variations we see in K2 peak separations in
the solar spectrum from the perspective of stellar activity studies 
where this parameter is correlated with the degree of chromospheric heating 
that is present.  In addition, we compare our Center Disk observations 
with the distribution of activity seen in samples of solar-type stars.

\subsubsection{K2 Peak Separations}
 
In the context of chromospheric physics, it is of interest to compare the 
behavior of the K2Violet vs K2Red peak separation in our temporal series of solar 
measurements (Fig. 6) with the predictions of the 
chromospheric scaling laws derived by \citet{ayr79}.  These scaling laws
relate chromospheric emission line profile characteristics, including
the Wilson-Bappu width, to fundamental 
stellar properties and the degree of mechanical heating in the chromosphere of 
a star.  Following eq. 15 in \citet{ayr79}; see also eq. 9 in 
\citet{lin79}, the scaling law corresponding to the K2 peak separation
can be written as 

%\begin{equation}
%$$
%log$~\Delta\lambda_{\rm K_2}~$=~ $-\frac{1}{4}$~(log~$\tilde{A}_{\rm Fe}~$+~log~$\tilde{g}$~+~5 log~$\tilde{T}_{eff}~$)~$-\frac{1}{4}$~log~$\tilde{F}$~+$\frac{1}{2}$~log~$\xi$~+~{\rm constant}~, 
%$$
%\end{equation}

\begin{equation}
$$
log$~\Delta\lambda_{\rm K_2}~$=~{\rm constant}+$\frac{1}{2}$~log~$\xi$~$-\frac{1}{4}$~log~$\tilde{F}$~$-\frac{1}{4}$~(log~$\tilde{A}_{\rm Fe}~$+~log
~$\tilde{g}$~+~5 log~$\tilde{T}_{eff}~$)~,
$$
\end{equation}

where $\Delta\lambda_{\rm K_2}$ is the separation of the the K2V and K2R peaks,
$\tilde{A}_{\rm Fe}~$ is the iron abundance, $\tilde{g}$ is the gravity, 
$\tilde{T}_{eff}~$ is the effective temperature, and 
$\tilde{F}$ is the chromospheric heating rate, each normalized 
to `quiet Sun' values.  The parameter $\xi$ is the chromospheric
Doppler width in velocity units.  The calcium abundance is assumed to scale
proportionally with the iron abundance.  The relative chromospheric heating 
rate, $\tilde{F}$, is measured by the chromospheric emission in the Ca~\footnotesize{II}\normalsize~
H and K lines. We note that in the case of the Sun the normalized 
logarithmic terms for metallicity and surface gravity in parentheses in 
eq. (1) are all identically zero.  We assume here that changes in 
effective temperature are negligible during the solar cycle. This assumption
is corroborated by our own results given herein for 
temperature sensitive lines as well as the lack of any observed changes in 
the solar radius that could be associated with 
the observed cycle variations in the bolometric irradiance.

Since we are mainly interested in
examining the variation of the solar K2 peak separations with the cyclic 
modulation of the chromospheric heating rate in the Sun-as-a-star, we write 
eq. (1) simply as 

\begin{equation}
$$

log~$\Delta\lambda_{\rm K_2}$~=~{\rm constant}~$-\frac{1}{4}$~log~$F^{\prime}(H+K)$~~, 

$$
\end{equation}

where $F^{\prime}(H+K)$ is the total net radiative cooling 
(i.e., chromospheric emission) in the Ca~\footnotesize{II}\normalsize~ H and K lines corrected
for the radiative equilibrium contribution; the Doppler 
velocity parameter  
$\xi$ along with unit conversions and the normalization factor for the 
chromospheric heating rate have been absorbed into the constant 
term.  In order to estimate $F^{\prime}(H+K)$ we convert our 
summed K Ca ~\footnotesize{II}\normalsize~ K and H index values to total
flux utilizing the scaling laws given in a stellar context 
by \citet{hal95} and \citet{hal96} as functions of $B-V$ color.  
For the Sun, we
adopt ($B-V$) = +0.65 \citep{van84}.  In order to account
for the non-chromospheric, photospheric contribution to the H and K lines, we
apply the empirical correction given by \citet{noy84}, again as
a function of $B-V$ color.  The result is
graphically displayed in Fig. 13 where we see a declining trend of the
solar K2 peak separations with increasing chromospheric heating as represented
by the increasing total chromospheric flux in the H and K lines of 
the Sun.  The decreasing
K2 separation with increasing chromospheric emission is in agreement with 
the trend predicted by the scaling laws derived by \citet{ayr79}.  The 
regression line has a slope of -0.22 $\pm$ 0.005, which is close to the
value of -0.25 for the slope in eq. (2).  We note, parenthetically, that
adopting colors for the Sun-as-a-star ranging from as blue as 0.62 to as
red as 0.67 does not change appreciably the above value for the slope of the 
best fit regression line.  

Besides the dependences on photospheric 
constants (e.g., metallicity, gravity, and effective temperature), 
the scaling law for the K2 peak separation in eq. (1) 
depends on the chromospheric Doppler width in addition
to the chromospheric heating rate.  We therefore suggest that
departures from eq. (2) above---and the cyclic variation of the 
Wilson-Bappu width (Fig. 6)---could be 
due to the cyclic modulation of the average chromospheric velocity field 
in the K2 peak formation region.
\subsubsection{Center Disk Ca ~\footnotesize{II}\normalsize~ K index Values
and Activity in Solar-Type Stars}

It is again of interest in a stellar context to compare our mean H and K index values 
for the Center Disk with that of solar-type stars, particularly stars that may be 
considered as Maunder-minimum candidates.  Our mean Center Disk value could be 
indicative of the values to expect in especially quiescent stars or even in 
the Sun during prolonged episodes of relatively reduced activity, as appears 
to have occurred during the Maunder-minimum period.

In order to compare our results with stellar data in the literature, we 
convert our summed Center Disk H+K index values to the Mt. Wilson S index 
using the empirical 
relations given by \citet{hal95} and \citet{hal96}.  The S index 
is based on the 
summed relative strengths in approximately 1 {\AA} bandpasses centered on the 
Ca~\footnotesize{II}\normalsize~ H and K line cores \citep{vau78}.  
We find the mean and rms for the S index 
corresponding to our center disk observations to be S = .133 $\pm$ 0.006.  This 
mean S index 
is at the lower limits of the distribution of chromospheric activity 
reported by \citet{hal04} (their Figs. 2 and 3 in their sample 
of 57 Sun-like stars).
  
In addition, our estimated mean HK index of 
150 $\pm$ 7 m{\AA} for the solar Center Disk data is about 
10\% less than the seasonal mean values, as measured over several seasons
of observation, for even the most quiet 
solar-type stars in the M67 
\citet{gia06}; their Fig. 3 in the solar-age and metallicity open 
cluster, M67, using an identically defined HK index.  There is a little more 
overlap between the mean HK index for the solar Center Disk data and the
M67 solar-type star data for individual seasons of observation.  In any
event, our mean center 
disk value in comparison with stellar observations appears to be representative 
of ``immaculate photospheres'' with little in the way of magnetic 
field-related nonradiative heating. 

\section{Results for Photospheric Lines}

In our archives this lowest level in the atmosphere is represented by the lines: Fe~\footnotesize{I}\normalsize~ 5379.579, C~\footnotesize{I}\normalsize~ 5380.323, and Ti~\footnotesize{II}\normalsize~ 5381.026\AA, see Fig. 1(a). C~\footnotesize{I}\normalsize~ 5380 is especially interesting because this 7.68 ev excitation line is formed deep in the photosphere close to the continuum in \citet{liv91}, p.1124. As previously noted, unlike other photospheric lines it weakens toward the limb and disappears there. In principle C~\footnotesize{I}\normalsize~ 5380 should be useful to determine temperature variations but the issue is complicated and the reader is referred to other papers on the subject \citep{liv82}, \citep{liv88}, \citep{pen06}.

In Fig. 14 we show the temporal record of the above lines near 5380\AA~corrected only for water blends. The initial period marked between (a) and (b) may be considered experimental where optical filters and slit widths were being selected. These data are not used. Epoch (b) to (c) is our most stable time series when no instrumental changes were made, and collimation was infrequent. These are the data studied by \citet{gra97a,gra97b}. The noise was sufficiently low that the annual variation of the apparent size of the solar disk projected on the smaller grating could be deduced from limb-effect variations. Point (c) in 1992 is where the larger grating was installed. This was followed by an interval of several years during which differing collimation procedures were tried. (The reader might wonder why it took such a long time to settle on such observational techniques. The answer is that any changes produced only small effects that were not evident until after several months of measurement). Finally, at point(d) such experimentation ceased and the system has remained unchanged  to the present (e).

Note that the strength of both the Fe and Ti lines increased with the larger grating at point (c). This is because the depth
of these lines increases near the limb. Limb regions were now better sampled by the new grating. Likewise, C weakened because this line is weaker near the limb. Why at point (d) the C intensity goes up, we do not know for certain. In fact all three lines show about the same increase. It could be that the intermediate slit width was made narrower as an experiment and this reduced grating scattered light (grating ghost and satellite structures). Another  obvious result is the increase of variance. As we mentioned, this is the consequence of changing gratings between different observing runs, there being two gratings now.  Alternating between IR and visible observations requires re-collimation  and introduces slight changes in spectrograph performance.

Another aspect seen in Fig. 14 is that many minor temporal artifacts are common to measurements in all three lines. By taking suitable line ratios we can eliminate most of these instrumental effects. The results are shown in Fig. 15. After this `correction' and a shift of the (d) to (e) mean level, owing to the limb effect, the record is flat within 0.18\% in the earlier epoch and 0.63\% in the recent. No activity cycle modulation or secular change is evident.

An exception to the above constant behavior is photospheric Mn~\footnotesize{I}\normalsize~ 5394.672\AA. This line is known to be sensitive to chromospheric conditions \citep{doy01,mal04}. Fig. 16 shows its temporal behavior and how it compares with Ca~\footnotesize{II}\normalsize~ K. The detailed explanation is controversial at present but may involve optical pumping from transition layer lines. Alternately, the explanation may lie, like Ca~\footnotesize{II}\normalsize~ H and K, in the surface distributions of faculae. In other words, the fine structure constant effect in this line makes it more sensitive to temperature \citep{dan06}.

\section{Results for High Photosphere Lines}

The upper photosphere is sensed by the cores of hundreds of medium to strong Fraunhofer lines. Mitchell compared FTS Full Disk spectra taken at the maximum of cycle 21 in 1980 with the following minimum in 1985 \citep{mit91}. He deduced an average increase of 1.4\% in central depth at minimum compared to maximum. Qualitatively this is what one would expect if the action of solar maximum hot faculae and plage is to slightly fill in line cores. Mitchell would have been surprised that his finding was not valid in the next cycle, but this is what we must now report.

Rather than study the hundreds of lines as Mitchell did, we turn our attention to selected unblended strong Fe~\footnotesize{I}\normalsize~ lines and the Na~\footnotesize{I}\normalsize~ D lines. Fig. 17 shows the FTS time history of these lines. Both Na~\footnotesize{I}\normalsize~ D and strong Fe~\footnotesize{I}\normalsize~ lines failed to respond to the minimum in 1996. Minimum cycle response in 1985 was 22\% for Na and 6\% for Fe, a significant amount. Fig. 18 indicates even more clearly the differences between what was observed and what was expected. It would be useful to have a standard line which did not vary as a reference check on FTS performance. The nearest we have to such a standard are the telluric oxygen lines near 6300\AA. These lines, although much narrower than solar lines, and therefore questionable as standards, indicate that the FTS performance was constant to 
within $\pm$2\%. We may note that Mn~\footnotesize{I}\normalsize~5394 does show a slight response in FTS records at solar minimum in 1996 (Fig. 16), but as explained in $\S$4 this behavior is compromised by optical pumping or some other mechanism. In any case the 1996 rise for this line is much less than in 1986.

We also observed using the spectrograph a group of strong lines in the UV near 3131\AA. See Figs. 1 (d) and 19. Their central depths are as follows: .90 for V~\footnotesize{II}\normalsize~ 3130.3; .76 Be~\footnotesize{II}\normalsize~ 3130.4;  .90 Ti~\footnotesize{II}\normalsize~ 3130.8; .65 Be~\footnotesize{II}\normalsize~ 3131.1; .60 Fe~\footnotesize{II}\normalsize~ 3131.7; .90 Cr~\footnotesize{II}\normalsize~ 3132.1. Because there is no continuum available for this region \citep{hou70},  we normalized to the local high point in the spectrum. Line blanketing is strong, and no systematic cycle variability is seen in these lines despite their apparent depths. The lack of variability in the relative intensities for these lines shows that the central intensity and the reference intensity vary by the same amount. This emphasizes the importance of using reference intensities formed well above the formation level for the line and as close to the photospheric continuum as possible.  

\section{Determination of Precipitable Water above Kitt Peak}
Many photospheric lines have water vapor blends. From the equivalent width of the telluric water line at 10832.2\AA, Fig. 1(c), we find for each observing day the precipitable water above Kitt Peak \citep{wal84}. This archive is displayed in Fig. 20 where a strong seasonal variation is evident with up to nearly 4 cm of water in the summer monsoon and down to 0.7 mm on the driest winter day. All data in this paper have been corrected for water where necessary. There are no discrete telluric absorption lines shortward of 4000\AA.
 
\section{Discussion}

We report here on 1215 single day observations of the Full Disk chromospheric Ca~\footnotesize{II}\normalsize~ K index from 1974-2006. Ca~\footnotesize{II}\normalsize~ K index tracked the 11 y activity cycle with peak-to-peak amplitude 
of about 25\% (Fig. 7).   
Ca~\footnotesize{II}\normalsize~ K3 intensity had a larger peak-to-peak amplitude of 37\%. Other chromospheric lines such as He~\footnotesize{I}\normalsize~ 10830, Ca~\footnotesize{II}\normalsize~ 8542, the CN band head at 3883\AA, and 
H$\alpha$ also followed the cycle but with reduced 
amplitudes (Fig. 11). Over the same time interval there were 243 Ca~\footnotesize{II}\normalsize~ K 2 arc-min resolution observations made at Center Disk. These were found not to be modulated by the cycle and constant within 1.4\%. This residual variance is presumed traceable to short lived, possibly cycle independent, ephemeral regions \citep{har93}. Similar observations in the low photospheric lines at 5380\AA~display no cycle modulation. We reported on this before and explained how we suppose that cycle magnetism threads its way between granules and does not affect them, i.e. heat them \citep{liv03,liv05}. High resolution ($\sim$2 arc-sec) scans of the quiet Sun yield a minimum Ca~\footnotesize{II}\normalsize~ K3 intensity of 0.04. This value represents the basal quiet non-magnetic Ca~\footnotesize{II}\normalsize~ K atmosphere and could represent conditions during the Maunder 
Minimum. Converted to stellar S, the {\it center disk} Sun resides at a 
position of minimum activity as found for the most quiescent solar-type stars.

The wavelength of the K3 feature is observed to vary in phase with the 
sunspot cycle for both Full Disk and Center Disk measurements. While the 
phenomenon is confined to Ca~\footnotesize{II}\normalsize~ K in our 
archives, in other solar-type stars the process could extend to other 
chromospheric or even photospheric lines (e.g., see \citep{hal98}), and might 
give rise to erroneous planetary detections.

The low photosphere is sensed by C~\footnotesize{I}\normalsize~ 5380\AA, and nearby Fe and Ti, of which there are 1154 observations 1976-2006. Compared with the chromospheric lines, their intensity variations are small and subject to instrumental effects. Particularly important was a grating change in 1992. After this date the pinhole image of the Sun did not over-fill the new grating as was the case earlier. This change altered how the global center-to-limb variation influenced the Full Disk measurements after 1992. We discussed these and other instrumental aspects and show how, by taking line ratios, we can correct for minor collimation effects. After correction, we have allowed our two main observing epochs to be brought into agreement with simple additive shifts. Following these corrections and shifts we demonstrate that the low photosphere has been constant to within 
$\pm$0.18\% in 
the early epoch and $\pm$0.63\% in the later (Fig. 14). No cycle modulation 
is detected. We conclude that the basal photosphere is not affected 
(i.e. heated) by cycle magnetism.

The high photosphere is sensed by the cores of strong Fe lines and the D lines of 
Na~\footnotesize{I}\normalsize~. These lines were observed from 
1980-2006 with the FTS. This instrument and the telescope did not change 
optically (except for mirror and beam-splitter coatings) over this 
interval as indicated by O$_{2}$~6303 (Fig. 18). Solar minimum in 1985 
produced a significant signal (dip) in the central depths of these lines 
(6\% in Fe~\footnotesize{I}\normalsize, 22\% in Na~\footnotesize{I}\normalsize). But during the following minimum these dips were not detected. We have no 
physical explanation for this behavior. The time series looks like it might 
be following a 22 y cycle, but more observations will be needed to 
resolve this anomaly.

In summary, what have 30 y of spectral observations told us about the 
physical Sun? Recall that the day-to-day fluctuation over a 3 month run on 
temperature sensitive C~\footnotesize{I}\normalsize~ 5380{\AA} yielded a 
fluctuation equivalent to 0.06\% in the solar constant \citep{liv77}. That 
behavior has basically continued over the entire observational period with 
no indication of cycle modulation or secular change. We conclude that the 
basal quiet photosphere is constant in temperature within our observational 
error as shown in Fig. 15. At the same time, and independently, chromospheric 
Ca~\footnotesize{II}\normalsize~ H \& K observed Full Disk has traced out the magnetic activity 
cycle and this has been shown by others to be a surrogate for UV 
emission. At Center Disk Ca~\footnotesize{II}\normalsize~ H \& K are constant, and so is the basal 
chromosphere. FTS observations suggest that the cores of high photosphere 
lines may vary with the 22 y Hale cycle. This has yet to be proven or 
understood. Perhaps another 20 y or so of observations will clarify this 
question.

\acknowledgments

A number of Chinese astronomers have contributed a year or more each to 
our program: Y-R Huang, Z. Liu, B. Ye, and Y. Wang. They did both programming 
and observing. O. Vince, Astronomical Observatory of Serbia, created the 
files for internet access. We thank the referee, David Soderblom, for his careful 
review of the original manuscript and for his suggestions that served to improve the
presentation.  The National Solar Observatory is operated by the 
Association of Universities for Research in Astronomy, Inc., under 
cooperative agreement with the National Science Foundation.

\appendix
\section{How to access and understand the data}

Both the original raw data, i.e. spectrum scans, and the reduced data in terms of measured line parameters are available. The URL for this is as follows: \\
http://diglib.nso.edu/cycle\_spectra.

{\it Raw Data:}
All spectrum scans come in pairs of odd and even records in fits format. These correspond to the forward and backward grating motions and should be combined without adjustment for the best S/N (hysteresis is removed at the telescope). Record lengths are 512, 1024, 2048, or 4096 points. Most are one sample per point (nspp), but some are two (as for Ca~\footnotesize{I}\normalsize~ 8542). The grating change in 1992 resulted in dispersion changes, and in some cases a necessary change of record length. A Fortran program at the data site (rescnew) allows one to rescale the new data to that of the old (dispersion and record length). Records that have been through the rescnew program may not have values at all points. A few early observations will differ from the above rules. Note that the data has not been normalized and all intensities are in AD units. This means that cloud problems can be recognized by a diminished ABSMAX (absolute maximum) in the record header.

{\it Reduced data:}
Observations have been reduced using a version of the program DECOMP. Normalization is to a 1, 2, or 3 point continuum
of specified values. The record is filtered, adjusted as needed in wavelength, and selected lines are measured for central depth (or central intensity) and equivalent width across a specified window (usually 140 m\AA). The results are placed in dat2.$\lambda\lambda\lambda\lambda$ files (e.g. dat2.5380). If corrections are made for water blends the dat2 files are converted to dat4 files. A typical file is in plain ASCII text:
mm/dd/yy MST nn r$_{i}$ ew$_{i}$ ....., where MST is Mountain Standard Time (the local time at Kitt Peak), nn is the number of records averaged, r$_{i}$ is the line depth and ew$_{i}$ is the equivalent width measurements for the i-th species. If a water correction is needed the correction value is given in the column header.

Reduced Ca~\footnotesize{II}\normalsize~ H and K archives are contained in fd.log2 and cd.log2 files, where fd and cd refer to Full Disk and Center Disk. Information for reading these files is contained at the site.

\clearpage

\begin{figure}
\includegraphics[angle=-90,scale=.65]{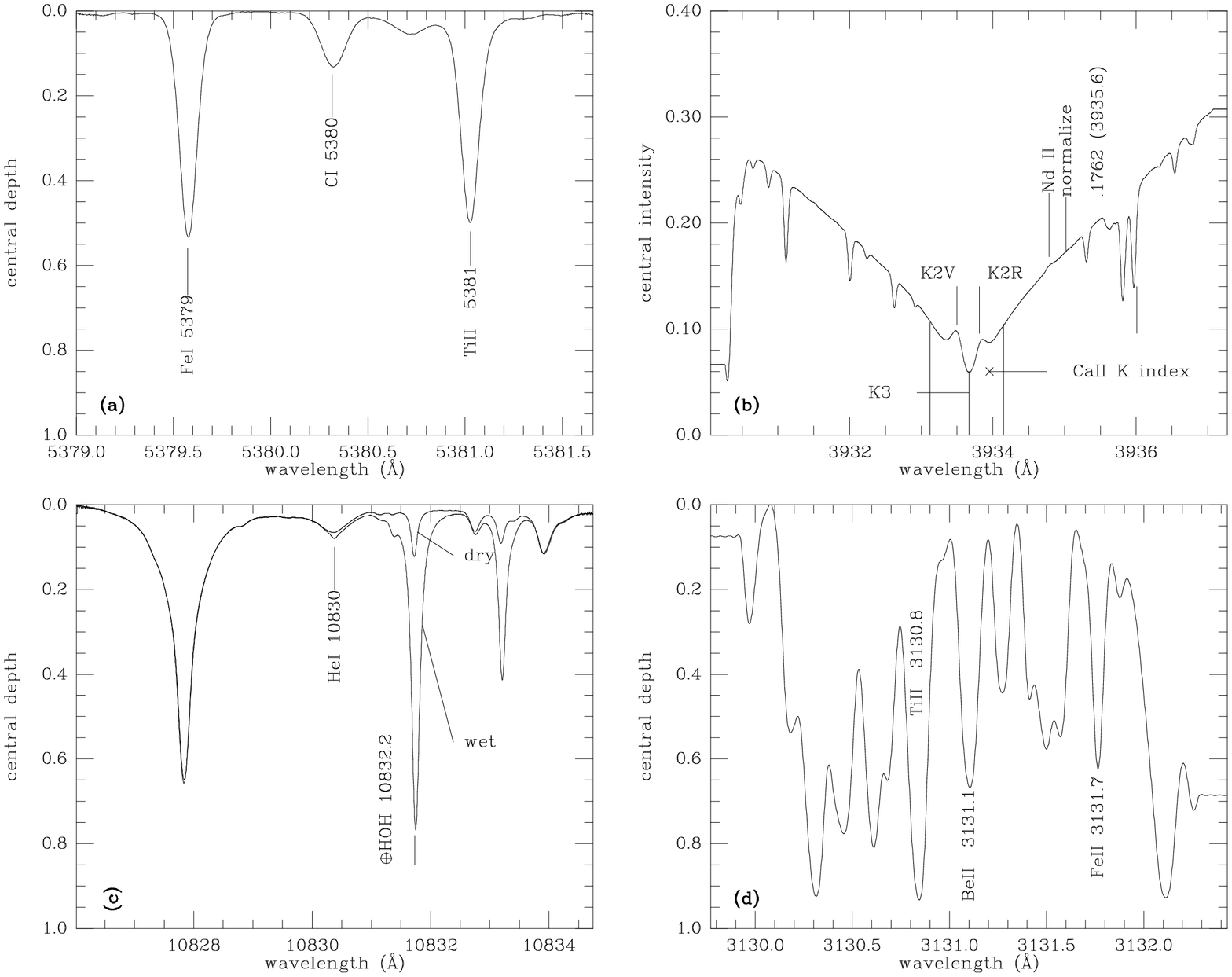}
\caption{Typical spectrum windows for this study. (a) C~\footnotesize{I}\normalsize~ 5380\AA~region; (b) Ca~\footnotesize{II}\normalsize~ K 3933\AA~region, at minimum phase, with normalization point indicated, and the K3 intensity, K index area, and other features shown. In the reductions the spectra are normalized to 0.1762 at the indicated position in the line wing; (c) He~\footnotesize{I}\normalsize~ 10830\AA~under wet and dry conditions. Note how the helium line blend with water becomes significant under wet conditions; (d) Be~\footnotesize{II}\normalsize~ 3131\AA~ region.}
\end{figure}
\begin{figure}
\epsscale{.50}
\plotone{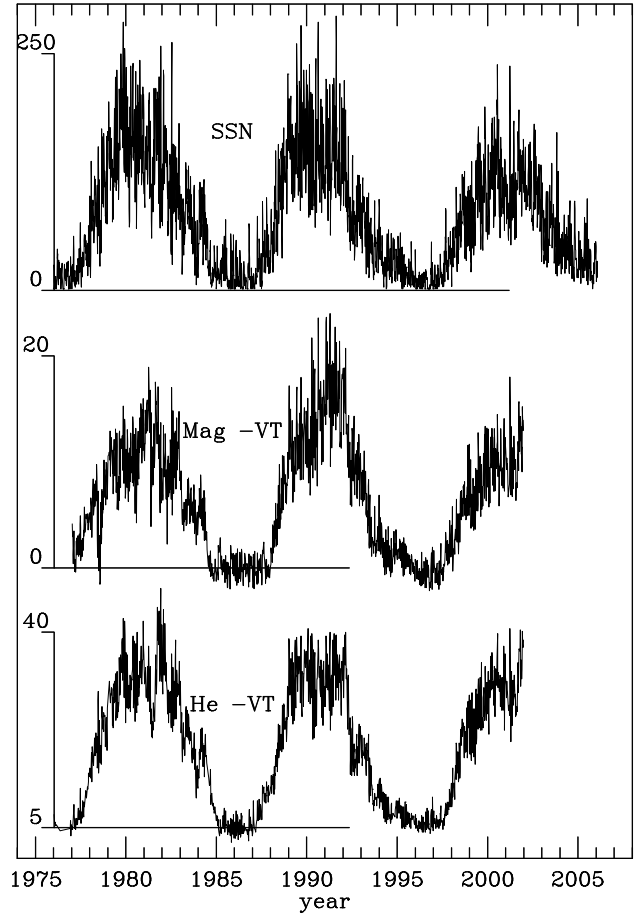}
\caption{Solar Full Disk indices. Top, sunspot number; Middle, total unsigned magnetic flux in Gauss as measured by the Kitt Peak vacuum telescope; Bottom, He~\footnotesize{I}\normalsize~ 10830 equivalent width in m\AA~summed (averaged) over the entire disk, also from the vacuum telescope.}
\end{figure}
\begin{figure}
\epsscale{.50}
\plotone{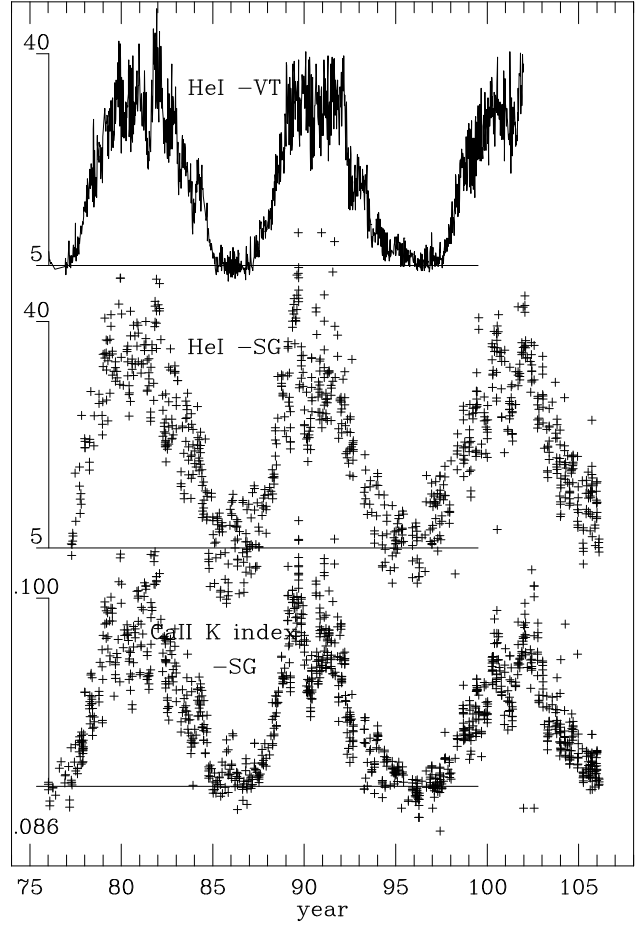}
\caption{Comparison of KPVT and McMath-Pierce spectrograph (SG) data. Top, summed He~\footnotesize{I}\normalsize~ 10830 equivalent width (same as Fig 1.); Middle, Full Disk He~\footnotesize{I}\normalsize~ 10830 equivalent width observed with the McMath-Pierce spectrograph; Bottom, Ca~\footnotesize{II}\normalsize~ K index also from the spectrograph.}
\end{figure}
\begin{figure}
\epsscale{.50}
\plotone{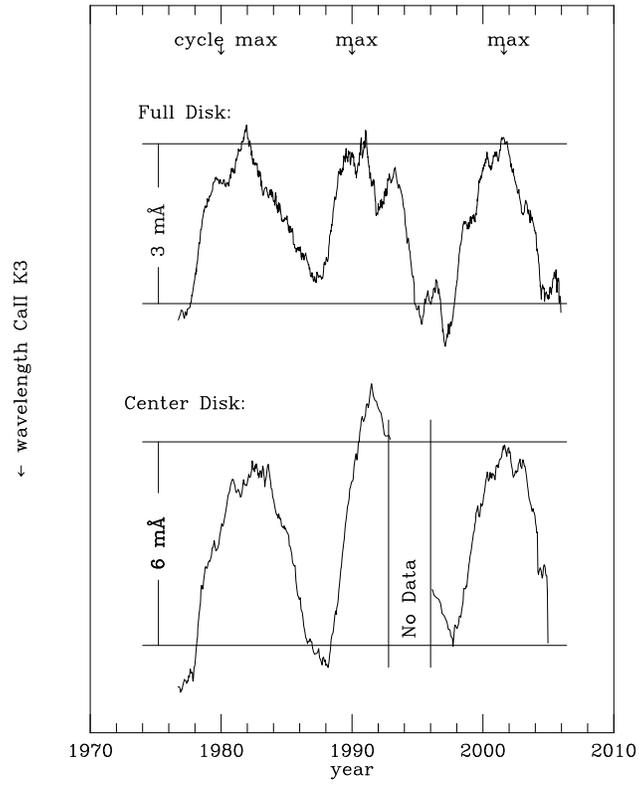}
\caption{Top, wavelength shift of K3 for Full Disk. Bottom, same temporal record for Center Disk. In the latter there was a 5 year hiatus of observations beginning in 1992.}
\end{figure}
\begin{figure}
\epsscale{.50}
\plotone{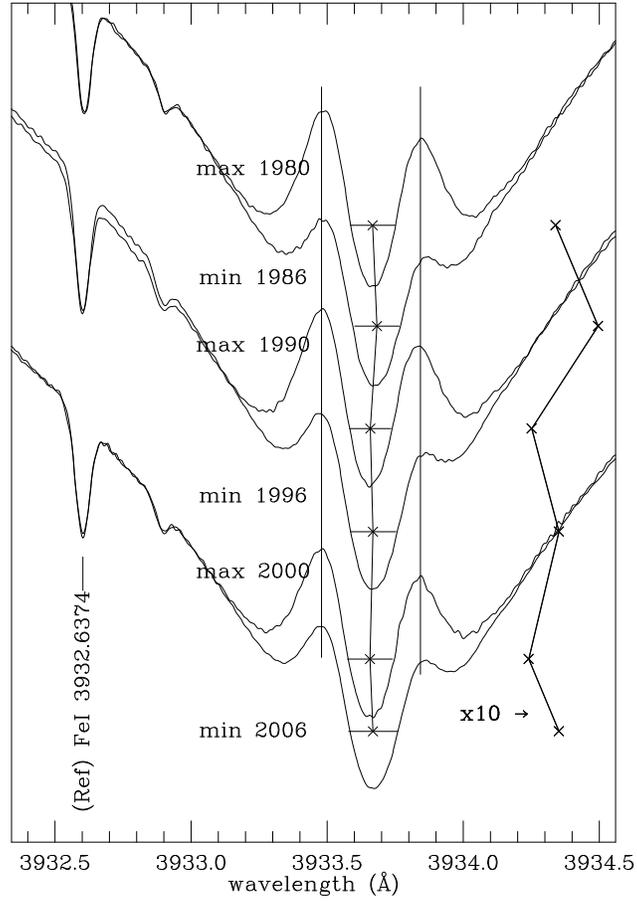}
\caption{Single day observations of the Ca~\footnotesize{II}\normalsize~ K232 feature at cycle minimum and maximum. The records have been shifted when necessary to keep the reference iron line at a fixed position. To the right is a 10-fold magnified plot of the K3 position.}
\end{figure}
\begin{figure}
\plotone{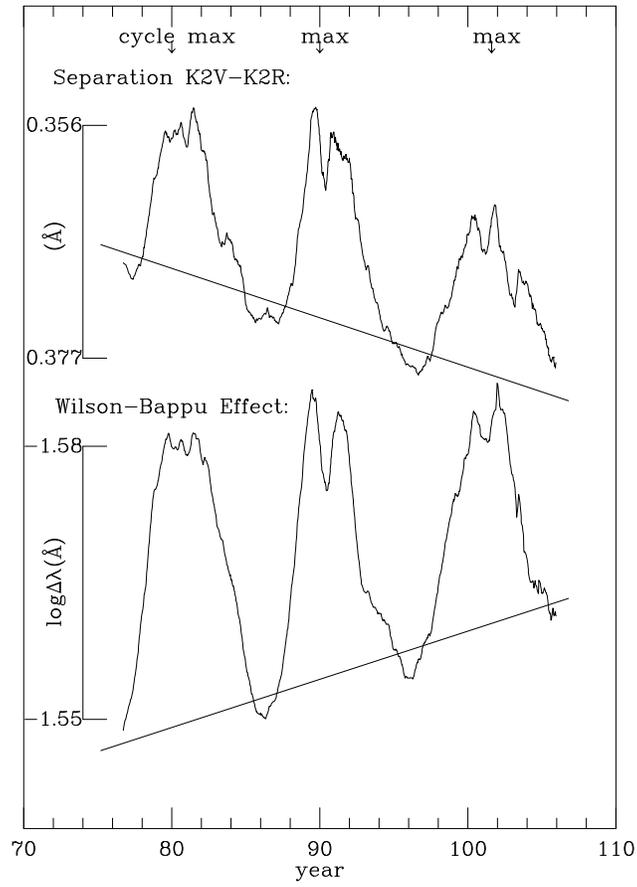}
\caption{Top, the wavelength separation between K2V and K2R with time. Wavelength increases down in the figure. Bottom, Wilson-Bappu effect( see text). Both parameters track the solar cycle as shown.}
\end{figure}
\begin{figure}
\plotone{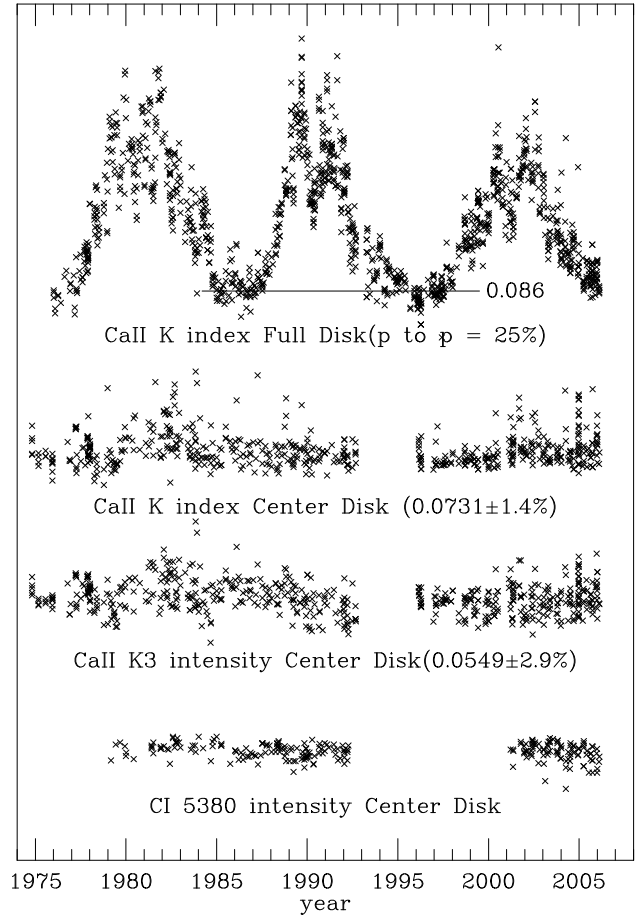}
\caption{Behavior of lines at Center Disk. Ca~\footnotesize{II}\normalsize~ K Full Disk index (for comparison); Center Disk: Ca~\footnotesize{II}\normalsize~ K index, with its mean value; Ca~\footnotesize{II}\normalsize~ K3 intensity, with mean shown; C~\footnotesize{I}\normalsize~ 5380 intensity}
\end{figure}
\begin{figure}
\plotone{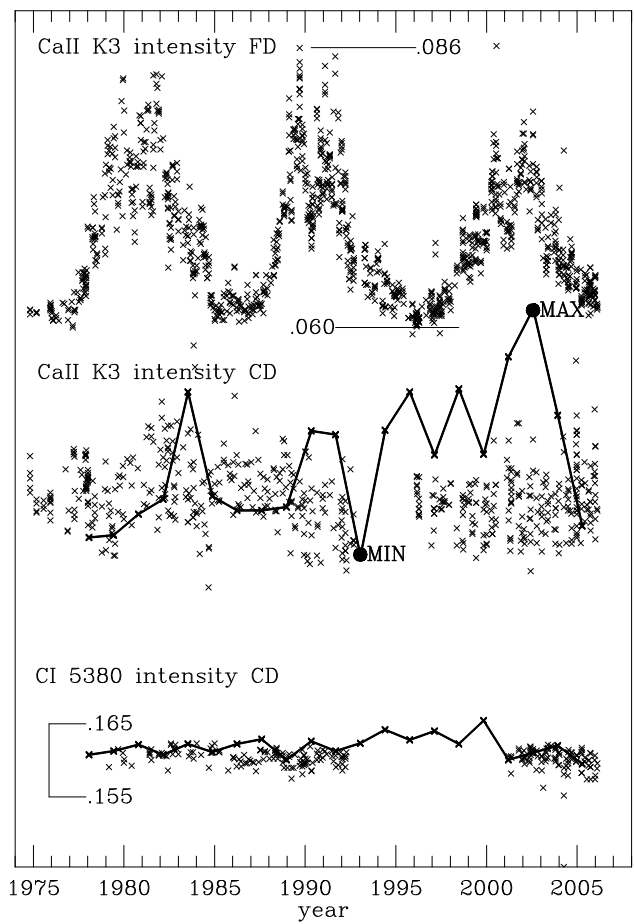}
\caption{Comparison of ranges of variability in time and over fine structure on the disk. Top; Full Disk Ca~\footnotesize{II}\normalsize~ K3 intensity 1976 to 2006 (for comparison). Cycle maximum peak-to-peak~=~(.086~$-$~.060)/.07~=~37\%. Middle; Center Disk: Ca~\footnotesize{II}\normalsize~ K3 intensity temporal record together (crosses), with K3 variation from a high resolution scan across a quiet region superposed. (The overlying plots are independent). Bottom; C~\footnotesize{I}\normalsize~ 5380 temporal record combined with a similar high resolution scan across (approximately) same area. The scan does not resolve granules which might be expected to show structure in C~\footnotesize{I}\normalsize~ 5380}
\end{figure}
\begin{figure}
\plotone{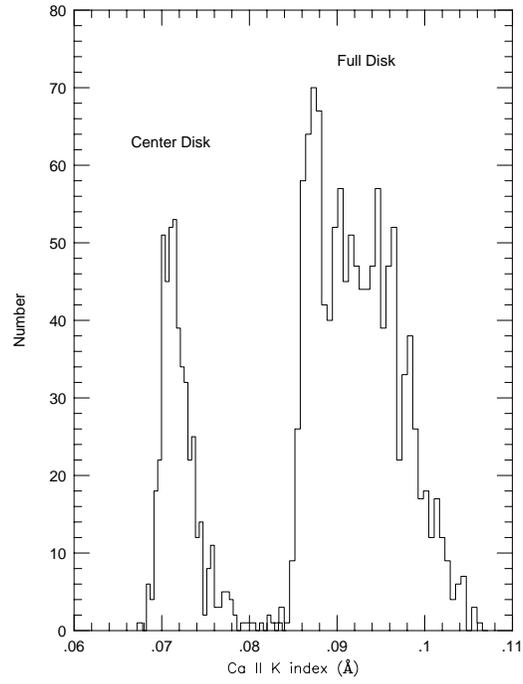}
\caption{Distributions of the Ca~\footnotesize{II}\normalsize~ K index values
observed for the Center Disk and Full Disk, respectively.  The distributions
are well separated with only minor overlap of the Center Disk K index 
measurements with the minimum values seen in the Full Disk observations}
\end{figure}
\begin{figure}
\plotone{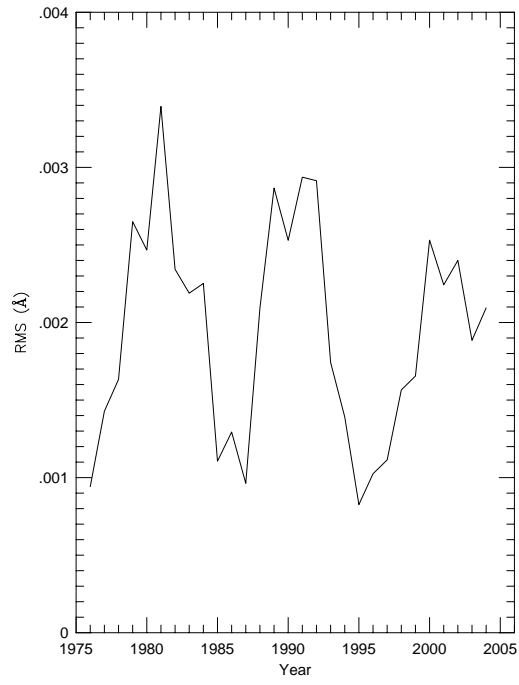}
\caption{RMS variation of the Ca~\footnotesize{II}\normalsize~ K index with 
one year bins. See $\S$3 for a discussion}
\end{figure}
\begin{figure}
\plotone{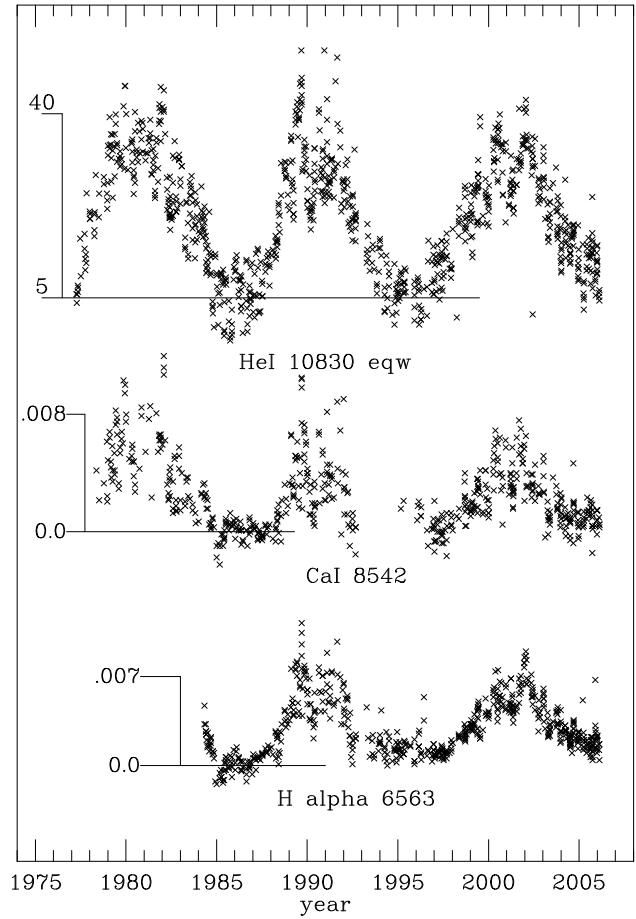}
\caption{Variability of Chromospheric Lines from Full Disk measurements using the 13.5-m McMath-Pierce spectrograph. Top, He~\footnotesize{I}\normalsize~ 10830 equivalent width; Middle, Ca~\footnotesize{II}\normalsize~ 8542 central depth (mean=.81); Bottom, H$\alpha$ central depth (mean=.83)}
\end{figure}
\begin{figure}
\plotone{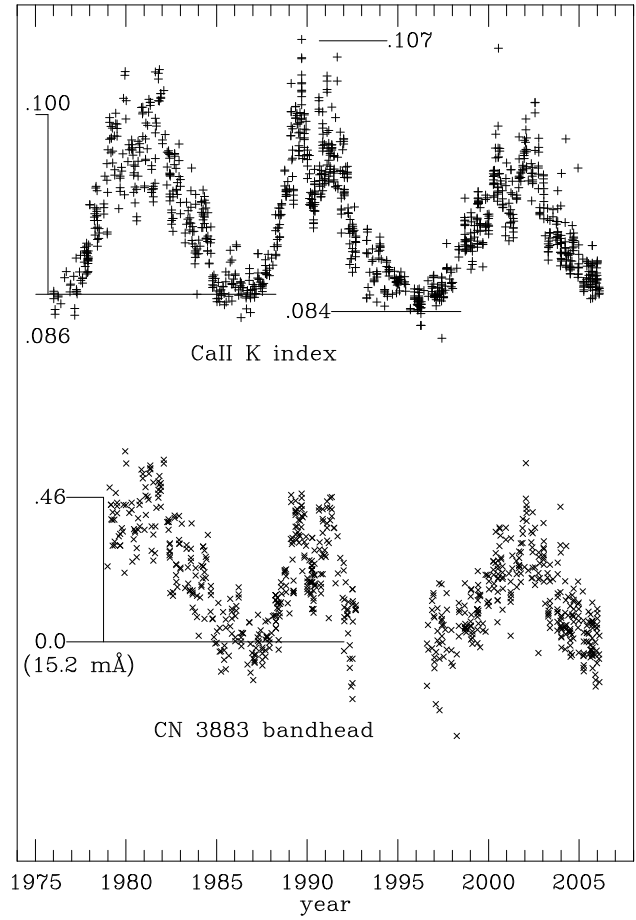}
\caption{Comparison of Ca~\footnotesize{II}\normalsize~ K index and CN band head index variability. Top; Ca~\footnotesize{II}\normalsize~ K index with nominal range indicated. (See Fig. 1 for definition of the Ca~\footnotesize{II}\normalsize~ K index). Cycle maximum peak-to-peak = (.107~-~.084)/.09~=~25\%~; Bottom, CN bandhead index at 3883\AA. The scale of the CN index variability of 0.0 to 0.46 m\AA~is the change above its nominal value of 15.2 m\AA~}
\end{figure}
\begin{figure}
\plotone{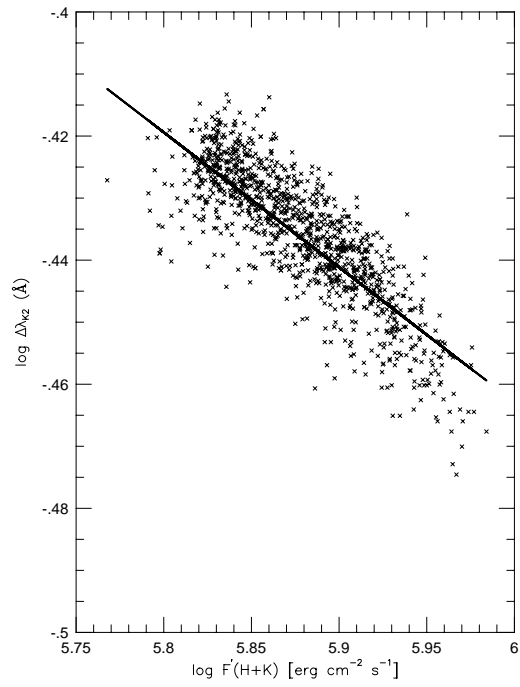}
\caption{The measured separations of the K2 peaks in the time series of 
solar spectra versus chromospheric H \& K emission as inferred from the 
Full Disk Ca~\footnotesize{II}\normalsize~ K index and equivalent H index 
data. The regression line has a slope of -0.22.  See $\S$3.1.1 for a discussion}
\end{figure}
\begin{figure}
\plotone{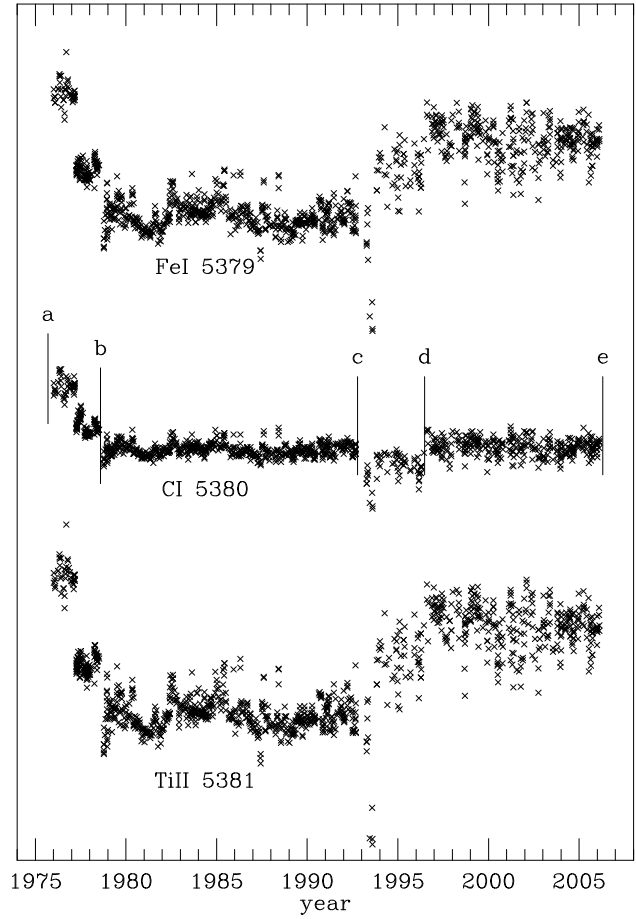}
\caption{Full Disk intensity of Fe~\footnotesize{I}\normalsize, C~\footnotesize{I}\normalsize, and Ti~\footnotesize{II}\normalsize~ lines. These are raw data corrected only for telluric water vapor lines. See text for the meaning of the marked epochs}
\end{figure}
\begin{figure}
\epsscale{.80}
\plotone{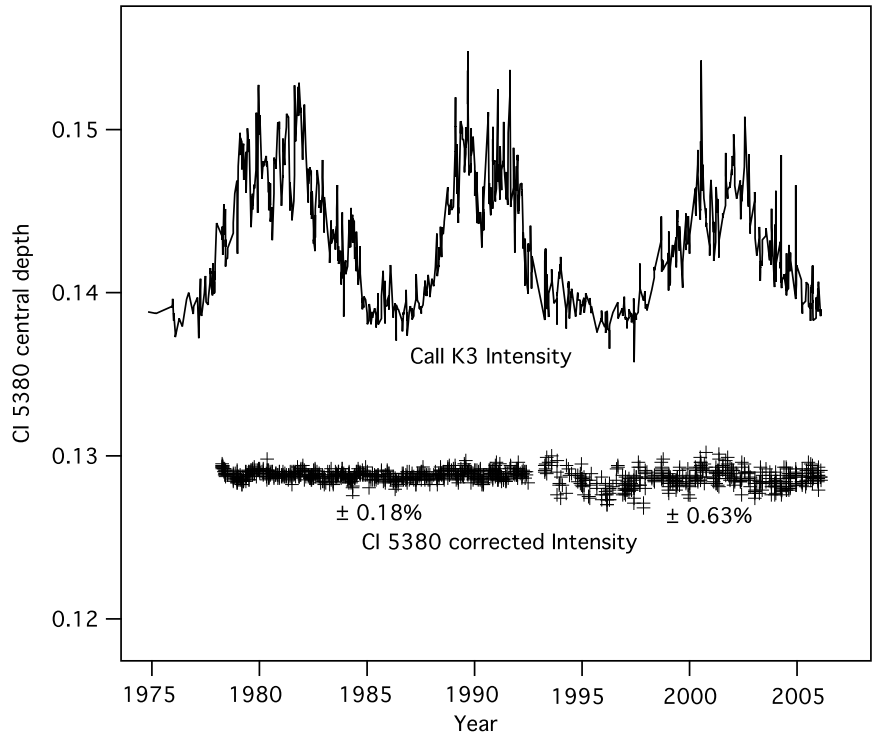}
\caption{Comparison of Variability of Full Disk Ca~\footnotesize{II}\normalsize~ K index (for cycle reference) and  C~\footnotesize{I}\normalsize~ 5380\AA~ line intensity corrected for instrumental effects using adjacent Fe~\footnotesize{I}\normalsize~ and Ti~\footnotesize{II}\normalsize~ lines}
\end{figure}
\begin{figure}
\plotone{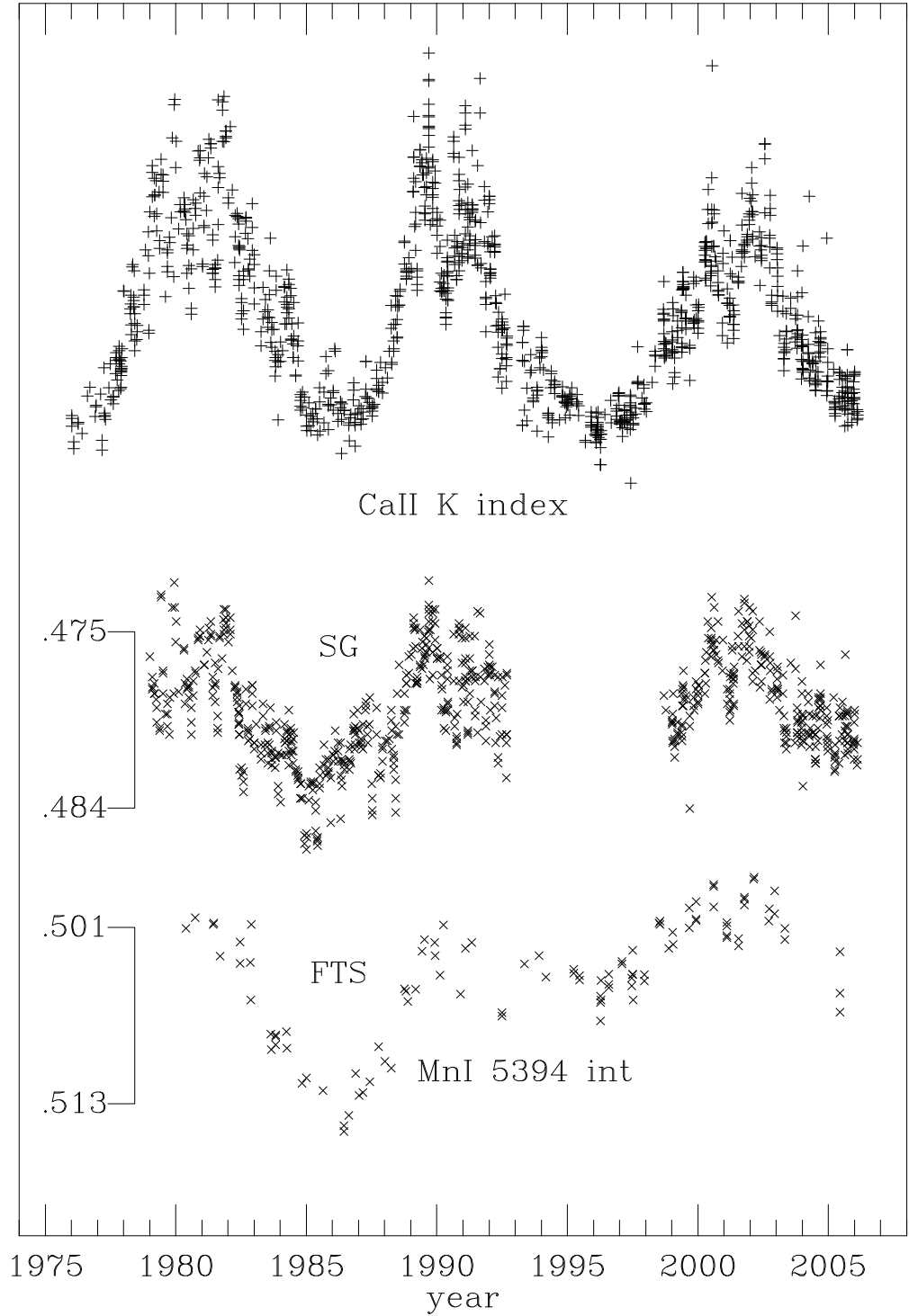}
\caption{Comparison of Variability of the Full Disk Ca~\footnotesize{II}\normalsize~ K index with the central intensity of the Mn~\footnotesize{I}\normalsize~ 5394\AA~ line from spectrograph and FTS data}
\end{figure}
\begin{figure}
\epsscale{.60}
\plotone{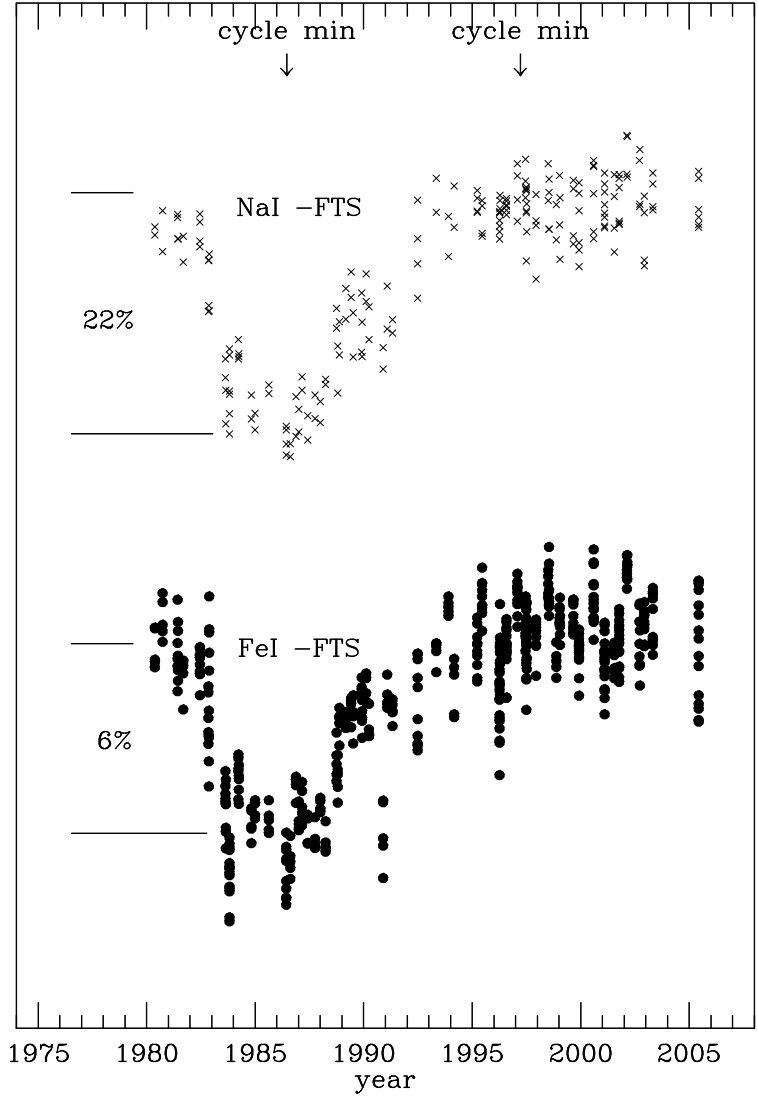}
\caption{Comparison of central intensities for Na~\footnotesize{I}\normalsize~ D lines and 5 strong Fe~\footnotesize{I}\normalsize~ lines: 5074.8, 5079.7, 5144.9, 5506.8, and 5586.8\AA~ from FTS data. Data for both sets of lines are scaled for minimum variance}
\end{figure} 
\begin{figure}
\plotone{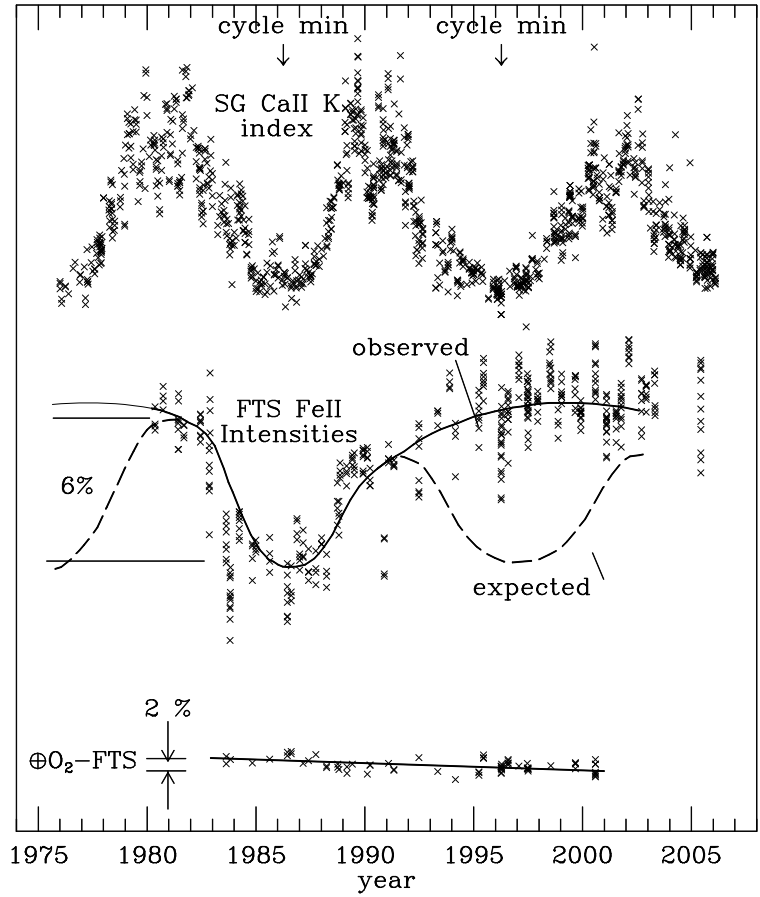}
\caption{Top: Full Disk Ca~\footnotesize{II}\normalsize~ K index; Middle, Central intensities of 5 strong Fe lines (see Fig. 17); Bottom, central depth of telluric O$_{2}$ 6302.00 and 6302.76\AA, adjusted to unity air mass. The solid lines fit the observed observations; the dashed represent what might be expected based on the response at the earlier 1986 minimum}
\end{figure}
\clearpage
\begin{figure}
\plotone{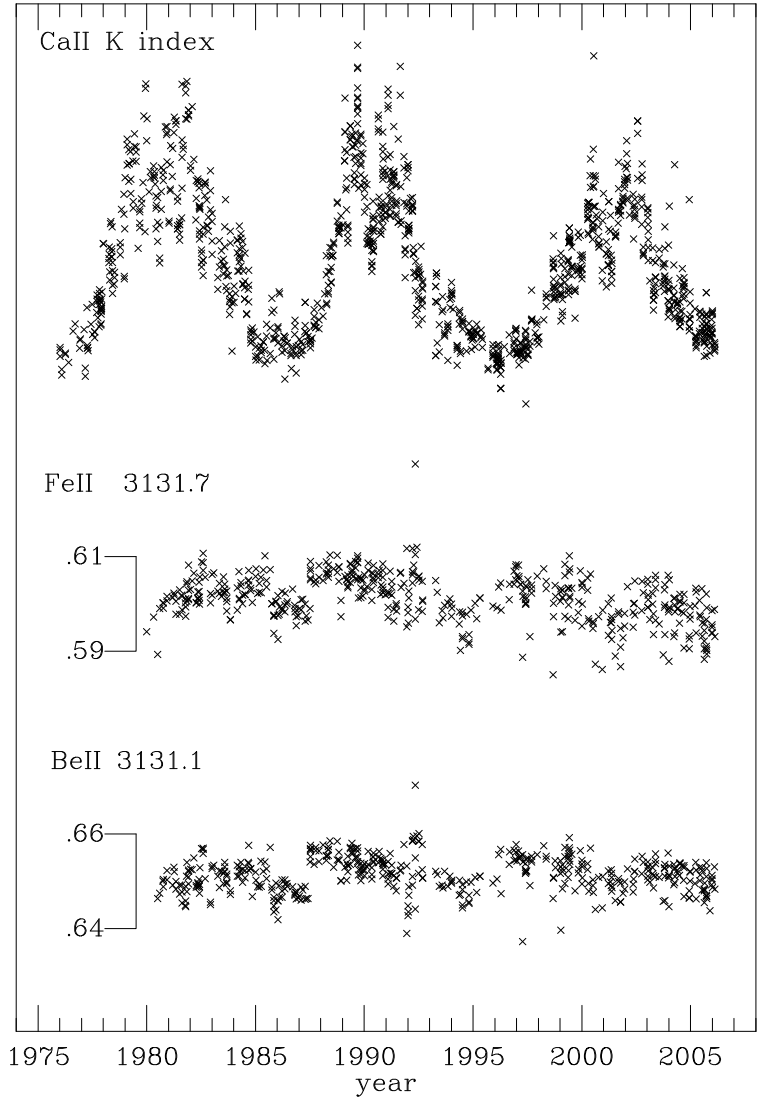}
\caption{Lack of solar cycle variability in central depths of UV Lines: Top, 
Full Disk Ca\footnotesize{II}\normalsize~ K index for comparison; Middle, 
central depth Fe~\footnotesize{II}\normalsize~ 3131.7; Bottom, 
Be~\footnotesize{II}\normalsize~ 3131.1\AA}
%\caption{This is the figure before using plotone f18.eps}
\end{figure}
\clearpage
\begin{figure}
\includegraphics*[scale=0.65,angle=-90]{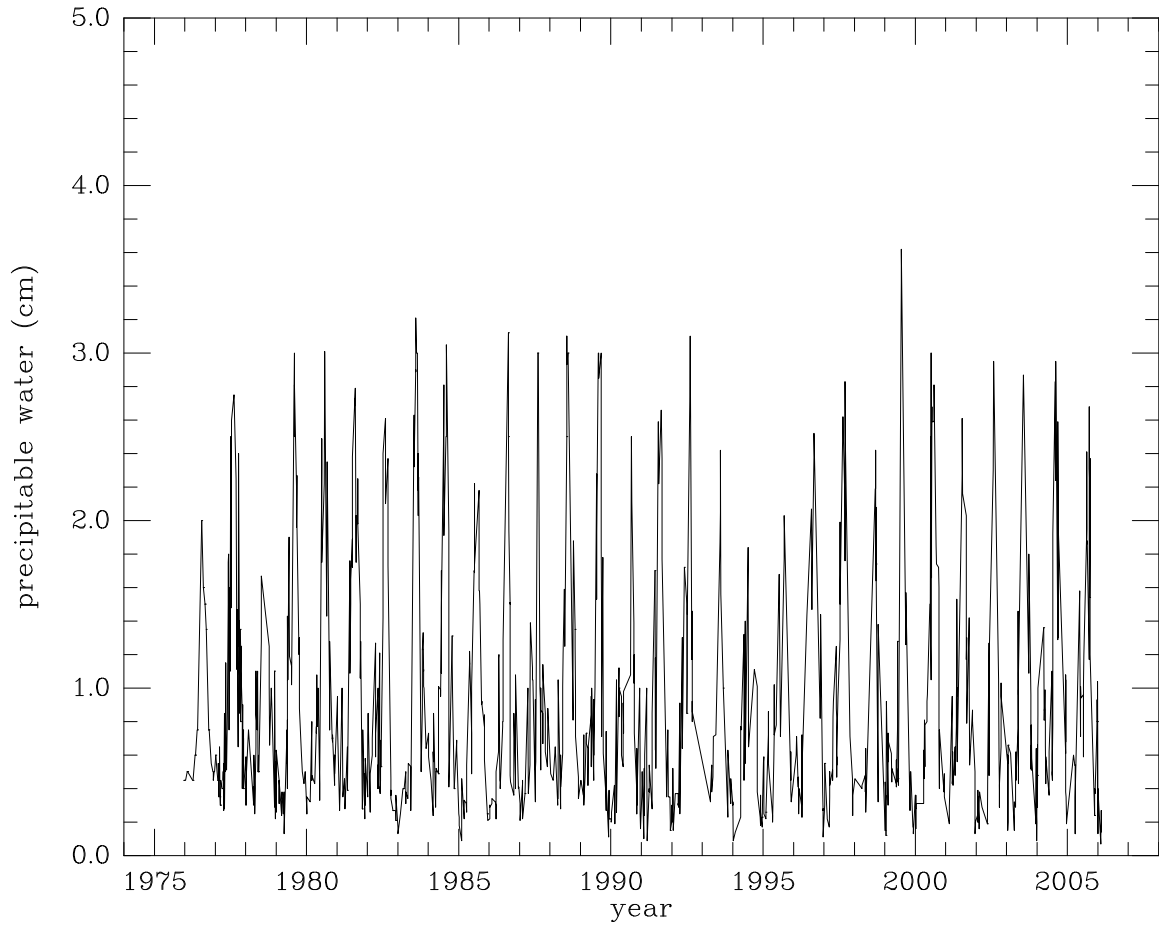}
%\plotone{f20.eps}
\caption{Record of Precipitable Water above Kitt Peak from 1976 to 2006}
\end{figure}


\begin{thebibliography}{}
\bibitem[Ayres(1979)]{ayr79} Ayres, T.R. 1979, \apj, 228, 509
\bibitem[Bailey \& Sheeley(1969)]{bai69} Bailey, W.L. \& Sheeley, N.R. Jr. 1969, Solar Phys., 7, 2
\bibitem[Brault et al.(1971)]{bra71} Brault, J.W., Slaughter, C.D., Pierce, A.K., \& Aikens, R.S. 1971, Solar Phys., 18, 366
\bibitem[Danilovic et al.(2006)]{dan06} Danilovic, S., Solanki, S.K., Livingston, W., Krivova, N., \& Vince, 2006, private correspondence
\bibitem[Donnelly(1988)] {don88} Donnelly, R.F. 1988, private correspondence
\bibitem[Doyle et al.(2001)]{doy01} Doyle, J.G., Jevremovic, D., Short, C.I., Hauschildt, P.H., Livingston, W., \& Vince, I. 2001, \aap, 369, L13
\bibitem[Engvold \& Halvorsen(1973)]{eng73} Engvold, O. \& Halvorsen, H.D. 1973, Solar Phys., 28, 23
\bibitem[Gray \& Livingston(1997a)]{gra97a} Gray, D.F. \& Livingston, W. 1997, \apj, 474, 798
\bibitem[Gray \& Livingston(1997b)]{gra97b} Gray, D.F. \& Livingston, W. 1997, \apj, 474, 802
\bibitem[Giampapa et al.(2006)]{gia06} Giampapa, M.S., Hall. J.C., Radick, R.R., \& Baliunas, S.L. 2006, \apj, 651, 444
\bibitem[Giovanelli \& Hall(1977)]{gio77} Giovanelli, R.G. \& Hall, D. 1977, Solar Phys., 52, 211
\bibitem[Hall(1996)]{hal96} Hall, J.C. 1996, \pasp, 108 313
\bibitem[Hall \& Lockwood(1995)]{hal95} Hall, J.C. \& Lockwood, G.W. 1995, \apj, 438, 404
\bibitem[Hall \& Lockwood(1998)]{hal98} Hall, J.C. \& Lockwood, G.W. 1998, \apj,
 493, 494
\bibitem[Hall \& Lockwood(2004)]{hal04} Hall. J.C. \& Lockwood. G.W. 2004, \apj, 614, 942
\bibitem[Harvey-Angle(1993)]{har93} Harvey-Angle, K.L. 1993, Magnetic Bipoles on the Sun, thesis, Univ. Utrecht
\bibitem[Houtgast(1970)]{hou70} Houtgast, J. 1970, Solar Phys., 15, 273
\bibitem[Jones \& Giovanelli(1983)]{jon83} Jones, H.P. \& Giovanelli, R.G. 1983, Solar Phys., 87, 37
\bibitem[Keller et al.(2003)]{kel03} Keller, C.U., Harvey, J.W., \& Giampapa, M.S. 2003, Innovative Telescopes and Instrumentation for Solar Astrophysics, Keil, S.L. ed., Proc. SPIE 4853
\bibitem[Livingston et al.(1977)]{liv77} Livingston, W., Milkey, R., \& Slaughter, C. 1977, \apj, 211 281
\bibitem[Linsky et al.(1979)]{lin79} Linsky, J.L., Worden, S.P., McClintock, W., \& Robertson, R.M. 1979, \apjs, 41, 47
\bibitem[Livingston \& Holweger(1982)]{liv82} Livingston, W. \& Holweger, H. 1982, \apj, 252, 375
\bibitem[Livingston \& Steffen(1988)]{liv88} Livingston, W. \& Steffen, M. 1988, Adv. Space Res., 8, no.7, 133
\bibitem[Livingston et al.(1991)]{liv91} Livingston, W., Donnelly. R.F., Grigoryev, V., Demidov, M.L., Lean, J., Steffen, M., White, O.R., Willson, R.L. 1991, p. 1124 in Solar Interior and Atmosphere, Univ. Arizona Press, Tucson
\bibitem[Livingston \& Wallace(2003)]{liv03} Livingston, W. \& Wallace, L. 2003, Solar Phys., 212, 227
\bibitem[Livingston et al.(2005)]{liv05} Livingston, W., Gray, D., Wallace, L., \& White, O.R. 2005, Large-Scale Structures and their Role in Solar Activity, ASP Conf Series 346, San Francisco
\bibitem[Malanushenko et al.(2004)]{mal04} Malanushenko, O., Jones, H., \& Livingston, W. 2004, IAU Symp 223, Stepanov, Benevolenskaya, \& Kosovichev
\bibitem[Mitchell \& Livingston(1991)]{mit91} Mitchell, W.E. Jr. \& Livingston, W. 1991, \apj, 372, 336
\bibitem[Noyes et al.(1984)]{noy84} Noyes, R.W., Hartmann, L.W., Baliunas, S.L., Duncan, D.K., \& Vaughan, A.H. 1984, \apj, 279, 763
\bibitem[Penza et al.(2006)]{pen06} Penza, V., Pietropaolo, E. \& Livingston, W. 2006, \aap, 454, 349
\bibitem[Pierce (1964)]{pie64} Pierce, A. K. 1964, \ao, 3, 1337
\bibitem[Sheeley(1969)]{she69} Sheeley, N.R., Jr. 1969, Solar Phys., 9, 347
\bibitem[Skumanich et al.(1984)]{sku84} Skumanich, A., Lean, J.L., White, O.R., \& Livingston, W. 1984, \apj, 282, 776
\bibitem[Title(1966)]{tit66} Title, A. 1966, Selected Spectroheliograms, Calif Inst. Tech, Pasadena
\bibitem[VandenBerg \& Bridges(1984)]{van84} VandenBerg, D.A. \& Bridges, T.J. 1984, \apj, 278, 679
\bibitem[Vaughan et al.(1978)]{vau78} Vaughan, A.H., Preston, G.W., \& Wilson, O.C. 1978, \pasp, 90, 267
\bibitem[Wallace \& Livingston(1984)]{wal84} Wallace, L. \& Livingston, W. 1984, \pasp, 96, 182
\bibitem[White \& Livingston(1981)]{whi81} White, O.R. \& Livingston, W. 1981, \apj, 249, 798
\bibitem[White et al.(1998)]{whi98} White, O.R., Livingston, W.C., Keil, S.L. \& Henry,T.W. 1998, Synoptic Solar Physics, ASP Conference Series Vol. 140, eds. K.S. Balasubramaniam, Jack Harvey, \& D. Rabin, p.293
\end{thebibliography}
\end{document}